# Scientific Objectives for UV/Visible Astrophysics Investigations:
## A Summary of Responses by the Community (2012)

*Paul A. Scowen[1], Mario R. Perez[2], Susan G. Neff[3], and Dominic J. Benford[3]*

**Abstract**

Following several recommendations presented by the Astrophysics Decadal Survey 2010 centered around the need to define *"a future ultraviolet-optical space capability,"* on 2012 May 25, NASA issued a Request for Information (RFI) seeking persuasive ultraviolet (UV) and visible wavelength astrophysics science investigations. The goal was to develop a cohesive and compelling set of science objectives that motivate and support the development of the next generation of ultraviolet/visible space astrophysics missions. Responses were due on 10 August 2012 when 34 submissions were received addressing a number of potential science drivers. A UV/visible Mission RFI Workshop was held on 2012 September 20 where each of these submissions was summarized and discussed in the context of each other. We present a scientific analysis of these submissions and presentations and the pursuant measurement capability needs, which could influence ultraviolet/visible technology development plans for the rest of this decade. We also describe the process and requirements leading to the inception of this community RFI, subsequent workshop and the expected evolution of these ideas and concepts for the remainder of this decade.

## I. Introduction

It has been recognized that at mid- and near-ultraviolet wavelengths ($90 < \lambda < 300$ nm), it is possible to detect and measure important astrophysical processes, which can shed light into the physical conditions of many environments of interest. For example, in the local interstellar medium (LISM) all but two (CaII H and K lines) of the key diagnostic of resonance lines are in the ultraviolet (Redfield 2006), which is depicted in Figure 1. In addition to the fruitful science areas that ultraviolet spectroscopy has contributed since the early 1970s, France et al. (2013a) have emphasized the role of ultraviolet photons in the photodissociation and photochemistry of $H_2O$ and $CO_2$ in terrestrial planet atmospheres, which can influence their atmospheric chemistry, and subsequently the habitability of Earth-like planets. However, only limited spectroscopic data are available for extrasolar planets and their host stars, especially in the case of M-type stars. Similarly, new areas of scientific interest are the detection and characterization of the hot gas between galaxies and the role of the intergalactic medium (IGM) in galaxy evolution (Shull et al 2012).

[1] Arizona State University, School of Earth & Space Exploration, P.O. Box 876004, Tempe, AZ 85287-6004, paul.scowen@asu.edu

[2] NASA Headquarters, Astrophysics Division, 300 E Street SW, Washington DC, 20546, mario.perez@nasa.gov

[3] NASA Goddard Space Flight Center, Cosmic Origins Program Office, Greenbelt, MD 20771, susan.g.neff@nasa.gov, dominic.benford@nasa.gov

NASA has invested substantial resources in advancing ultraviolet imaging and spectroscopy in space missions, most recently, in the Small Explorer Galaxy Evolution Explorer (GALEX; Martin, 2011) and the Hubble Space Telescope (HST) instruments Cosmic Origins Spectrograph (COS; Green et al. 2012) and the Space Telescope Imaging Spectrograph (STIS; Woodgate et al. 1998). Legacy NASA missions include Copernicus, IUE, Astro I/II (UIT, WUPPE, HUT), ORFEUS (SPAS I & II; IMAPS, BEFS, TUES), EUVE, FUSE, HST instruments (FOC, WPFC2, FOS, GHRS, ACS) and some limited imaging capabilities in SWIFT. Other substantial investments include technology development and maturation in devices such as detectors, optics, coatings, and other supporting technologies like spectral calibrators, coronagraphs, and occulters. Similarly, NASA has an active technology program in the area of suborbital platforms, including stratospheric balloons and sounding rockets and several of these payload experiments include ultraviolet imaging and spectroscopy capabilities (e.g., FIREBALL (Milliard et al. 2010), FORTIS (McCandliss et al. 2004), FIRE (Gantner et al. 2011), IMAGE (Gibson et al. 2000), PICTURE (Mendillo et al. 2012), SLICE (France et al 2013b)). The scientific achievements by all these missions and experiments include broad areas of interest, however, they will not be reviewed in this paper.

**II. RFI Motivation**

A detailed analysis supporting the Astrophysics Decadal Survey 2010 "New Worlds, New Horizons" (NRC/NAS 2010a, p.21) recommendation that NASA "carry out a mission-definition program" indicates some underlying desired aspects for the future of an ultraviolet/visible space capability. Foremost, under the small scale space recommended activities (ibid, Table ES.1, p.5) this language is included: *"Technology development benefiting a future UV telescope to study hot gas between galaxies, the interstellar medium, and exoplanets,"* with a target goal of spending $40 million dollars during the decade.

Similarly, as part of the main report, further elaboration of these arguments is presented (ibid, p. 220): *"Key advances could be made with a telescope with a 4-m diameter aperture with large FOV and fitted with high-efficiency UV and optical cameras/spectrographs operating at shorter wavelengths than HST."* As part of the panel reports (NRC/NAS 2010b, p.296), in particular of the Panel on Electromagnetic Observations from Space, additional justification of these recommendations is described: *"It is difficult to imagine an advanced UV-optical telescope that would not, in addition to spectroscopic capability, include high-resolution cameras, probably with considerably greater areal coverage than has been possible with the Hubble."* They find that the combined promise of greatly improved ultraviolet diagnostics of galaxy evolution, capable high-resolution wide-field UV/visible general astronomy capability, and studying exoplanets for habitability and life merit a mission of large scope. Warnings of potential pitfalls are also included in the language of this report, in that a large UV-visible telescope would be *"a facility that cannot happen without significant technology advances to make these science goals attainable and – especially if a large aperture is required – affordable."* Finally, the actual recommendation is focused on some actionable options*: "The panel views this as a unique opportunity that requires a **dedicated technology development program.** Because it believes that a UV-[visible] telescope is a particularly strong candidate for a new start in the 2021-2030 decade, the panel recommends pursuit of several different technologies over the coming decade…"*.

In summary, the Decadal Survey 2010 clearly recommended as aspects of the "mission-definition program" that NASA support technology development and conduct mission concept studies leading to a future ultraviolet/visible observatory dedicated to measuring the interstellar medium and the hot gas between galaxies, studying the evolution of normal stars, galaxies and planetary systems, and surveying exoplanets for habitability and the potential for life.

**III. RFI Inception**

The Astrophysics Division (ApD) at NASA Headquarters is the implementing entity of the space activity recommendations issued by the Decadal Survey 2010. In this role, ApD has issued an *"Astrophysics Implementation Plan"* (2012) articulating a near-term response (2013-2017) to these recommendations considering the contemporary set of policy and budgetary constraints, which were different from the ones assumed when the formulation of these recommendations was envisioned.

Furthermore, as part of the tactical advice from the science community via the Astrophysics Subcommittee (ApS), which is part of the NASA Advisory Council (NAC), the ApS started three Program Analysis Groups (PAGs). These groups, one for each of the three themes within the ApD, namely, the Physics of the Cosmos PAG (PhysPAG), Cosmic Origins PAG (COPAG), and the Exoplanet Exploration PAG (ExoPAG), have contributed to technology formulations and prioritizations, which have already influenced NASA solicitation for proposals and technology maturation plans for this decade.

In the spirit of responding to the Decadal Survey 2010 recommendation of defining the future UV/visible space astrophysics capability, ApD, in coordination with the Cosmic Origins Program Office, issued a Request for Information (RFI) on 2012 May 25 soliciting community input for compelling science drivers that could be accomplished by observations with an ultraviolet/visible space telescope. While the Decadal Survey specifically mentioned a 4-meter-diameter aperture, the expected implementation of a UV/visible mission is far enough in the future and the anticipated capability needs sufficiently imprecise that specifying a particular mission configuration seemed premature. The RFI was intended to elicit a representative (although not comprehensive) set of science investigations from which a well-defined, suitably detailed, and appropriately derived set of science requirements could be developed. The fundamental mission parameters for the future UV/visible mission would be derived only after such a set of science requirements is defined. The RFI represented the initiation of a larger process of defining those requirements by delineating the science investigations such a mission might ultimately fulfill. The RFI therefore invited all interested parties to address the question "What observing proposal will you write in the next decade?" in order to engender a mission that might realize the capabilities implicit in their response.

This RFI is part of an ApD plan to start a broad outreach effort to engage the overall community to present and discuss new and persuasive science drivers that could be part of future mission concepts and mission enabling technologies. During the remainder of this decade, a technology development plan is starting to emerge that could prioritize technology needs and support and downselect technology implementations to be part of compelling and mature mission concepts to be considered by the next Decadal Survey. Future workshops and multiple interactions with the community are being planned in order to achieve these goals.

Several aspects of the UV/visible RFI were designed with the idea of broad outreach in mind. The Cosmic Origins Program Office held virtual question and answer sessions to clarify the RFI to elicit a broad range of science investigations. Responses were allowed from any person or group, as often as desired, provided that the submission would represent a clear idea of the detailed science investigation the responding group would hope to conduct. The science investigations were specifically to be conceived of without the restrictions of any prior mission implementation concepts, to enable any new ideas to be considered with equal weight. Even though when the Cosmic Origins Program conducted this RFI, the science investigations solicited were illimitable in scope, supporting COR, PCOS, ExEP, and even non-ApD goals. No restrictions were placed on the size, complexity, or cost of the mission required to address the proposed science investigations. Finally, the RFI results were to be presented and discussed publicly in a Workshop in 2012 September (Section IV), to promote the demonstrations as part of a larger process that would define the science requirements for a future mission definition.

## IV  RFI Process and Responses

### IV. a. RFI Process

The RFI was released on 2012 May 25. It was announced to the community through NASA Acquisition Internet Service (NAIS), Federal Business Opportunities (FedBizOpps.gov), NSPIRES, the AAS mail exploder, the COR email exploder, and the COR program website. Two question and answer (Q&A) sessions were held for community interaction, on 2012 June 5 and on 2012 July 17. A frequently asked questions (FAQ) file was generated from email inquiries and from the Q&A sessions, and were posted on the COR website. Submissions were made through NSPIRES. The RFI closed for responses on 2012 August 10. Proprietary or limited-distribution investigations were not accepted; all proposed investigations were to be made public.

Submissions were required to address, at least in part, goals of the Cosmic Origins (COR) program ("Explore the origin and evolution of the galaxies, stars, and planets that make up our universe"). Investigators were asked to restrict their proposed investigations to those that could be accomplished in the UV and visible parts of the electromagnetic spectrum, covering the wavelength range approximately 90-1000 nm. Extensions to longer or shorter wavelengths were not forbidden, but were expected to be supplemental to the UV/visible band emphasis. The responses were intended to provide a working set of science requirements supporting future UV/visible wavelength investigations.

**IV. b. Responses**

Thirty-four individual responses were submitted to the UV/visible Astrophysics Investigations RFI. Most responses were submitted on behalf of larger teams of co-respondents. A list of the responses is given in Table 1, and an interactive list, with links to the investigations, may be found at
http://cor.gsfc.nasa.gov/RFI2012/rfi2012-submissions.php.

The RFI invited responses from all interested parties, regardless of nationality. Twenty-nine responses were from US investigators (from 13 states) and 5 were from outside the US (France, Canada, Spain, and the United Kingdom). Principal Investigator (PI) institution types included US and non-US Universities, NASA Centers, and the Space Telescope Science Institute (STScI). A total of 219 individuals participated in the responses, representing 21 states and 11 countries, and employed by Universities, National Research Centers/Laboratories, Research Institutions, and Observatories.

Responses were grouped into several primary science areas, as shown in Table 1: Stars and Stellar Evolution, Nearby Galaxies, Intergalactic / Circumgalactic Medium, Galaxy Evolution, Planets / Solar System (including exoplanets), and active galactic nuclei (AGN). Several responses addressed multiple science areas, and the responses included a rich diversity of more specialized topics.

Responders were asked to provide science-driven capability requirements for their investigations, including:
- Imaging / Spectroscopy / Time Domain
- Field(s) of View
- Physical / Angular Resolution (both desired and minimum requirement)
- Spectral Resolution, if relevant (desired and required)
- Wavelength range(s) (lower and upper limits / desired and required)
- Sensitivity (desired / required)
- Dynamic range (desired / required)
- Other requirements

As an ensemble, these science-driven requirements span a wide range of angular resolution (microarcseconds to near arcminute), Field-of-View (<10′ × 10′ to nearly 1° × 1°), wavelength coverage (90 nm to 1000 nm), and spectral resolution ($R \equiv \lambda/\Delta\lambda \sim 100$ to 100000+). There were 7 responses requiring photometry only, 13 requiring spectroscopy only, and 14 requiring both. Ten responses required spectroscopic multiplexing; seven were for Multi-object spectrographs (MOS), two were for Integral Field Units (IFUs), and one was for a slitless wide-field spectral imager. The responses included eight investigations in the time domain, two interferometers, two coronagraphs, and two polarimeters.

The science-driven requirements summarized in Table 1 are discussed in more detail in Section V. By way of simplistic analogy to prior missions, most of the responses appear to fall into three main groups:

i) *Short wavelength UV instrument ("super FUSE"):* these investigations focused mostly on far-UV spectroscopy; were optimized for λ≈ 91–125nm; requested a wide range in spectral resolution ($R \sim 6,000$–$100,000$); desired but did not always require spectral multiplexing (a Multi-Object Spectrograph (MOS) or Integral-Field Unit (IFU)); and typically indicated that angular resolution was not always a driver.

ii) *Wide-field imager ("super GALEX"):* these investigations requested a wide field of view with an angular resolution usually ~0.1″, where spectral resolution was not usually a driver.

iii) *Multiplexing high spectral resolution spectrograph ("super COS"):* these investigations generally sought a MOS or IFU with 10+ object multiplexing; a spectral resolution $R=20,000$-$100,000$; a wider range of desired wavelength coverage ($\lambda_{short} \sim 91.2 - 300$ nm, $\lambda_{long} \sim 300$-$1700$nm); and a typical angular resolution of ~1″.

All respondents were invited to present their investigations at a workshop on September 18, 2013, at the Space Telescope Science Institute (STScI) in Baltimore, MD, USA, and to participate in a follow-up meeting by the Cosmic Origins Program Analysis Group (COPAG) on September 19 also at the STScI.

**V. RFI Workshop**

The Cosmic Origins Program Office coordinated a UV/visible Mission RFI Workshop on 2012 September 18 at the STScI. The workshop served as a venue to discuss, present and categorize the 34 science investigation submissions. For organizational convenience, the short (five minute) presentations were grouped by primary science goal into five categories. These are: the intergalactic medium, star formation and nearby galaxies; stars; and broader science topics. On average, seven presentations in each category were made. We summarize the science cases presented in the same order.

In order to provide some commonality of science requirements, we elected to use this subdivision into broad subject areas to consider the range of capabilities that were common between the approaches advocated. However, with such a broad response there would naturally be submissions that stand clearly alone and represent singular visions of what the future of UV/visible science could look like. In this section, we attempt to do the submissions justice in discussing their overarching goals, but we also strive to define a series of common drivers and capabilities that could deliver the majority of the science proposed.

**V. a. Intergalactic Medium**

Several proposals were centered on the use of the large number of diagnostic lines in the far-ultraviolet (FUV; 91.2-120 nm) to open doors to information about the nature of material in the Intergalactic Medium (IGM), and what can we learn about the cycle of material between star formation in galaxies and how the resulting material is ejected from those galaxies and subsequently returned via other mechanisms.

One example of this approach was a proposal that intended to use QSOs as backlights to allow absorption line studies of the intervening IGM. The innovative approach, other than pushing hard into the FUV, is to use this methodology to trace invisible baryons – the material that has not participated in star formation or in galactic assembly and is therefore left out of most assays of material between the galaxies. Specific challenges, other than the need to optimize both throughput and detector quantum efficiency in the near-ultraviolet (NUV; 120 to 360 nm) and FUV is that many QSOs are simply too faint for some of the probing observations needed. However, the promise is too great to give up on the approach and advances in technology could yield dramatic dividends. This work would require deep FUV high-resolution ($R>20,000$) spectroscopic capability. A second proposal focused on similar goals by proposing to trace and conduct a census of baryonic matter in the low redshift universe in the various phases of the IGM. This latter work would require deep NUV/FUV high-resolution spectroscopic surveys using a dedicated 6-8m aperture UV mission. The spectrograph would need very high throughput ($A_{eff} > 3\times10^4$ cm$^2$) and high spectral resolution ($R > 40,000$). Another proposal also used a similar approach but used AGNs as the backlit sources to determine how BHs accrete matter and grow over time. This approach of using AGN as backlights for IGM, circumgalactic medium (CGM) and interstellar medium (ISM) studies opens the door for innovative approaches such as reverberation mapping of the broad line region (BLR) around AGNs, quantifying the outflow of material, and the degree of radiation reprocessing that occurs. Such work requires very high-resolution angular and spectral FUV spectroscopy. The critical factor here is the necessary time domain capability requiring spectroscopy with time resolution ~1000s, with a field of view of only <1″, but with an angular resolution of ~10 mas. This work requires a spectral resolution of $R>40,000$ over a waveband of 91.2-320 nm with a sensitivity $5\times10^{-16}$ ergs/cm$^2$/s in 2000s.

Using similar diagnostics in both the FUV and EUV, another proposal concentrated on the evolution of the background ionization field threading the space between galaxies. The main issue here of course is the nature of the problem of reionization – how did this cosmological event get energized and how long did it take? A strong suspect is Lyman-continuum and Lyman-α photons that "leak" from star forming galaxies to ionize the IGM. The premise of the proposal is that in the FUV there are a relatively small number of corrections necessary to derive the galactic UV luminosity function in the redshift range $0<z<3$. A second proposal was centered on essentially the same goals but used lensing magnification of distant galaxies to provide higher sensitivity tracing and probing of the high redshift universe to establish the Lyman-continuum energy budget. Such work would require both FUV wide-field imaging and spectroscopy using 10× HST sensitivity at <300 nm, using detectors with lower read noise and better charge transfer efficiency for imaging at 100nm, coupled with 3-10× the field of view of HST-WFC3, and delivering $R$>5000 below the Lyman limit.

Taking an alternative approach, and instead of mapping FUV absorption signatures, another proposal outlined a way to use IGM/CGM emission mapping to probe baryonic structure formation across cosmic time. The idea is to detect and characterize the IGM emission to determine the physical properties of the IGM and trace baryonic structure formation using that same emission. Such work requires the use of FUV multi-object spectroscopy over modest sized fields. Specific capabilities proposed include an IFU with a 4×4 arcmin$^2$ field of view, a MOS instrument with a 20×20 arcmin$^2$ field of view, and 1-5″ resolution, delivering $R$~1000-5000, at $\lambda$~100-400 nm covering 0.05<$z$<1.5. The ultimate goal is the achievement of a sensitivity of 100-5000 photons cm$^{-2}$ s$^{-1}$ sr$^{-1}$ for the CGM, and 5-100 photons cm$^{-2}$ s$^{-1}$ sr$^{-1}$ for the IGM.

**V. b. Star Formation in Nearby Galaxies**

Stepping a little closer, there were several proposals submitted that focused on what could be learned about star formation as a global process by considering entire galactic systems or by conducting large local surveys of star formation apparent in our own Galaxy to provide statistical support for global trends in star and planet formation modes as observed in a variety of environments.

Starting with our own Galaxy, one proposal centered on conducting a wide field UV/visible imaging survey of massive star forming complexes to understand the formation mechanisms and survival rates of the star formation process in such environments. The argument is that the majority of stars form in such environments because of the initial mass function, and as such these environments offer a key insight into the rate at which stars and planets can form in close proximity to massive stars. Such a program requires the delivery of wide field (>10′, >200 arcmin$^2$) UV/visible imaging at diffraction limited (0.04″ at 300nm) resolution behind a 1.5m-4m aperture with a large suite of optical filters. Such a program requires an efficient well-corrected optical design married with high yield efficient detectors (to allow the production of a large number of detectors to tile a focal plane).

Another proposal hit on a theme that shows up in several different places in this summary – that of understanding how material, and in particular the chemical elements, are distributed and dispersed into the CGM and the IGM. Of particular interest is how baryonic matter flows from the IGM into galaxies and from there into stars and planets. This proposal was rooted in conducting high-resolution multiband UVOIR survey imaging of the Magellanic Clouds (MCs), together with a narrowband survey of H II regions and the diffuse warm ISM in the Clouds. A complementary FUV spectroscopic survey of 1300 early-type stars would provide direct tracers of both the ISM conditions as well as providing insight into the nature of these stellar atmospheres to provide checks for atmospheric codes. As above, this program required a large field of view (>200 arcmin$^2$) with diffraction limited imaging at 300nm behind an aperture of 1.5-4m with a large suite of filters. The FUV spectroscopy requires next generation reflective coatings combined with new microchannel plate (MCP) technology and a spectral resolution of R>30,000.

Along similar lines, another proposal focused on the use of massive stars as a tool to measure the range of properties stars can be formed with: the range of mass, composition, convection, mass-loss, rotation rate, binarity, magnetic fields, and ultimately cluster mass affect massive stars and their feedback mechanisms. Such a wide-ranging survey requires UV spectroscopy of a statistically significant sample of OB stars in the MCs, together with UV/visible imaging and spectroscopy of large sample of local SF galaxies to study the escape rate of Lyα photons, since these have a direct impact on the ecology of the CGM and have implications for cosmological questions such as reionization. This program focused on a waveband of 120-160nm, with UV spectroscopy and UV/visible imaging over a field of view of 25″×25″ at an angular resolution of <0.1″. The spectroscopy required a spectral resolution equivalent to the COS-G130M mode, but also required an aperture >10m to allow the resolution of O stars in 1 Zw 18.

A much larger-scope investigation centered on analysis of the conditions for life in the local universe. At issue is the question of how cosmic feedback affects habitability – specific measurements include a better understanding of the physics of hot atmospheres, and the changes in interplanetary environment that could be caused by a range of feedback mechanisms. Such an ambitious program requires considerably larger samples of targets than are currently available, and to perform ultra-high FUV spectroscopy and wide field FUV imaging for those targets. The targets are too far apart to be addressed by multi-object spectroscopy, and so require another design solution. Overall the program requires better sensitivity in mirrors, coatings, detectors, gratings and filters. The required spectroscopy needs $R>100,000$ and a large effective area to get to the faint levels needed, combined with high resolution imaging across a waveband of 100-300nm.

Taking a more global view, another program focused on deriving entire star formation (SF) histories for nearby galaxies. This would require the photometry of resolved stellar populations in those nearby galaxies to directly measure their SF histories, as well as separating components and structures within each galaxy. The program would assemble larger galaxy groups, and include ultra-faint dwarf galaxies, to provide an order of magnitude increase in sample size and/or the limiting signal since the spatial volume sampled scales as the aperture cubed. This investigation would require an 8m telescope to perform wide-field UVO imaging at the diffraction limit at 500nm across wide-field high-resolution imaging fields not available from the ground (or the UV).

Using local galaxies as cosmological analogs, one proposal performed UV/visible wide-field imaging and spectroscopy to perform so-called "near field cosmology". Such an approach uses globular clusters to provide a fossil record of the earlier SF era – but requires the ability to measure globular cluster (GC) properties at a high level of accuracy to provide access to the outer halo SF history for a particular galaxy. Such a program required wide-field diffraction limited UVOIR imaging: a field of view ~1000 arcmin$^2$ at an angular resolution of ~0.05″. Spectroscopically a spectral resolution of $R$~3000 is needed across a waveband of 150 nm – 5 microns at a sensitivity level capable of delivering a S/N=30 for V=22.5 in 10 hrs (i.e. 8m of aperture).

A final proposal in this section intended to use the best performance available in both imaging resolution and sensitivity to extend the state-of-the-art for galaxy evolution studies. The intent is to understand how the diverse array of present-day galaxies came to be – to understand how does SF proceeds in different environments. The program uses spectral energy distribution fitting in the UV to split the degeneracies in reddening and temperature using UV-bright hot stars. The program requires wide-field UVO imaging across a field of view of ~30′ with an angular resolution of ~0.007″ across a waveband of 110-600nm.

**V. c. Stars**

Proposals focusing on stellar physics tended to concentrate on the new insight that could be gained by adding aperture or extremely high resolution spectroscopic measurements to the mix, as well as polarimetry and extremely high angular resolution imaging from formation flying imaging systems that could perform optical interferometry.

Massive stars are recognized as being important objects to understand their formation and evolution mechanisms since they deposit so much energy and material back into the ISM and affect star formation propagation. In that regard, one proposal outlined how it is important to understand how molecules and dust form in the interacting winds of massive stars. Such a project requires the use of a larger aperture visible-IR telescope with imaging spectroscopy. In addition, such work would also require the use of better 3-D models of such wind-wind interactions. This work required the delivery of imaging resolution of 0.01″ coupled with spectroscopy at $R$=10,000, time domain sampling, over fields of view <2″, with spectroscopic angular resolution ~0.005″, with both delivered over the visible-IR.

It is recognized that stellar atmospheres can serve as repositories of fossil records of the ISM from which they formed, and as such we can use them as tracers and an imprinted historical record of ISM heavy element evolution. Another proposal laid out that the necessary detection and measurement of heavy elements in stellar atmospheres requires high resolution FUV spectroscopy over wide fields, because of the wealth of diagnostic lines from a wide variety of species over the FUV band (see above). This requirement also implies the need for improved FUV reflective coatings as well as larger apertures to offset the losses due to coating losses as well as interstellar absorption. This work required access to 190-310nm, but with spectroscopic resolution of $R$~60,000, and at 10 times better sensitivity than HST-STIS, all delivered over a field of view ~10′×10′.

An especially innovative area of interest involves the study of stellar magnetospheres, winds, activity and their environments for a variety of stellar types over extensive periods of time. This analysis would provide direct insight into the evolution of magnetic fields in stars and about their internal dynamos. It would also help to explain how the interaction and shaping of the solar system and conditions for formation and evolution of planetary systems all hinge on better understanding of the magnetic fields threaded through a star's structure. Such insight requires UV and visible spectroscopy, as well as UV and visible spectropolarimetry, the latter of which is quite challenging from an instrumental design perspective, and what requirements it places on an optical design. This ambitious proposal required wavelength access of 117-870nm, with spectroscopic resolution in the UV of $R$~2000-100,000, and in the optical of $R$~35,000-80,000, but also required a dedicated mission duration of 4-12 years to achieve its goals. A second proposal was very similar to this idea, but included time series capabilities to greatly improve the resolution of rates of change and the agility with which a system can change its configuration and how such changes can affect the stellar environment.

The most challenging topic proposed involved dramatically improving our insight into mass transport processes and their role in the formation, structure, and evolution of stars and stellar systems. The kinds of angular resolution necessary to provide a dramatic advance in this field require the development of sub-mas level angular resolutions – to provide UV/visible spectral imaging that is capable of resolving stellar surfaces and environments. Such a capability requires the use of large diameter (0.5-1.0km) sparse aperture telescopes flown in formation to perform interferometry. This very innovative approach required a field of view of only ~4×4 mas, with an angular resolution ~0.1mas, and a spectral resolution of ~1nm across 120-660nm ($R$~300).

**V. d. Galaxy Evolution**

Many of the proposals in this category had marked similarities to the IGM proposals in terms of approach and science, but were sufficiently different to be discussed separately. But the similarities served to underline that there were an emerging set of capabilities that could service multiple avenues of research.

Using high time resolution high cadence (once a day for 180 days) mapping of UV emission from the gas around AGNs allows the construction of UV velocity-delay maps to track the flow of the high-ionization gas. This data can be used to construct reverberation maps of the BLR around those AGNs. Along a similar vein, the inner structure of AGNs can be mapped using UV/visible imaging at sub-mas levels. This knowledge can then be used to investigate and assess the role of such AGNs in the process of galactic formation and evolution. These kinds of observations can only be achieved with a space-based long-baseline (0.5-1.0 km) interferometer (UVOI). Such an instrument would require a field of view of only 4×4 mas, with an angular resolution of 0.1 mas, a spectral resolution ~1 nm over 120-300 nm. Desired sensitivities would be as low as $5\times10^{-14}$ ergs/cm$^2$/s for lines such as CIV (155nm).

Returning to the use of extragalactic Lyman-α observations, Lyman-α can be used to probe the lowest mass galaxies, the cosmic web, dark clouds, Pop III stars – all at UV wavelengths (150-360nm) using a dedicated survey telescope with a field of view >0.1 deg$^2$. The observations needed would have to be taken using slitless spectroscopy at $R$=100 and 5000.

Looking at the more global topic of galactic assembly and SMBH/AGN growth, wide-field imaging surveys can be conducted to assess how did galaxies evolved from the very first systems to the types we observe nearby. Because the objects at $z>7$ are very faint and very rare the observations require wide-field imaging combined with diffraction-limited optics to build enough of a sample that is statistically significant. What is of interest here is the evolution of the faint-end slope of the dwarf galaxy luminosity function, and what it can tell us about tracing the reionization history using such Ly-α emitters. Such work would require UVOIR observations using a large suite of filters behind an aperture of at least 2.4m.

Another proposal focused more on a dedicated UVOIR spectroscopic all-sky survey with the specific intention of understanding galaxy evolution – to understand how galaxies evolved to form the Hubble sequence we observe locally today and to establish which processes were responsible. The proposed survey consists of a 0.2-1.7 micron spectroscopic survey of $10^6$ galaxies at $z>0.8$.

**V. e. Broader Science**

Many proposals submitted did not conform to any of these common research areas and as such need to be mentioned on their own standalone merits and what capabilities they call out.

A large proposal was made to fly a dedicated flagship-class mission that concentrates on the exoplanet science that can be done of nearby stars while co-existing with a UV/visible astrophysics mission. Such a proposal could characterize planetary systems and their formation mechanisms, while conducting UV/visible detection and characterization of rocky planets. Such a flagship mission would require an aperture of 4m diameter or larger and use an internal coronagraph and external starshade in tandem. It was pointed out that some similar work could be done with a 1-2m mission but would only provide access to Jovian-class systems.

Provision of the same kinds of FUV/NUV capabilities discussed above could also open up capabilities that could be used to study the inner regions of protoplanetary disks (<10 AU) over planet formation timescales of ~$10^6$-$10^7$ yrs. A fundamental understanding of gas disk lifetimes and their structure determine how planets form their gas envelopes and therefore determine final architecture of exoplanet systems. This work would require FUV/NUV MOS/echelle spectroscopy. Specifically MOS spectroscopy over 120-180nm using a field of view ~10-20′ at $R$~3000 with a sensitivity < $10^{-15}$ ergs/cm$^2$/s/nm in 1000s, and echelle spectroscopy over 100-180nm at $R$~150,000 with a sensitivity < $10^{-16}$ ergs/cm$^2$/s/nm in 10,000s. The low-resolution NUV spectroscopy over 170-400nm could be done at $R$~100.

In a more general manner, a broad proposal laid out the Solar System science objectives that could be possible with a next generation UV/visible space observatory. The proposal identified a sequence of science drivers that would provide a local reference point for considerations of the origin and evolution of stars and planetary systems we see elsewhere in our local Galaxy. Such capabilities would require the provision of moving target tracking, as well as the ability to resolve time variable phenomena. General capabilities included UV/visible imaging and spectroscopy. The imaging field of view could be around 50″, with angular resolutions of ~0.03-0.05″, and spectroscopically a spectral resolution of $R$~100-10,000 would be sufficient.

Another broad-reaching proposal was to map and track the metallic evolution of the IGM, and through that data the physics and contents of galactic haloes, the evolution of UV irradiated environments and ultimately the emergence of life. Such work requires a very large number of spectroscopic lines of sight, coupled with narrow band UV imaging and spectroscopy, a large collecting area, large photon-counting detectors, improvements in coatings and UV optics materials. The proposal envisioned an UV survey of the Galactic plane that would be enabled by a fundamental improvement in our knowledge of molecular transitions lines. The required capabilities are broad involving a field of view that can range from ~1′-1°, angular resolutions that can range ~0.001-1″ with resolutions $R$~40-1000 over ~90-320 nm.

A broad proposal was brought forward for a wide-field UVO survey telescope with 0.15″ resolution – to provide a wide field imager for high resolution surveys that could support dark energy (DE) / dark matter (DM) science. The concept would provide for some PI programs and would include slitless spectroscopy. The (Canadian) proposal was made looking for partnerships within the US to start Phase A studies, and continue work on the optical design. The current design reported is an off-axis TMA with imaging bands in the UV, $u$ and $g$ only, with a field of view of ~1.2×0.6 deg$^2$, an angular resolution of ~0.15″ across ~150-550nm using 700 Mpix cameras behind a 1m aperture.

An impassioned second pitch was made for the large number of diagnostic lines available in the EUV/FUV that can enable unique astrophysics in the Lyman UV. The use of this waveband could provide access to the CGM and enlarge the available target sample. Such work cannot be done with an aperture less than 8m in size but would allow additional science such as determination of chemical abundances in star forming galaxies, the effect of UV on exoplanet biosignatures, and of course (see above) the nature of reionization and the escape fraction of ionizing radiation from star forming galaxies. Necessary resolutions range as high as $R$>30,000 with a sensitivity 10× better than HST-COS and a 10-100× MOS capability with improved FUV coatings.

## VI. Analysis of Submissions

The range of science drivers laid out in the 34 submissions received under the RFI ran the gamut of science driven by star and planet formation all the way up to drivers rooted in a fundamental understanding of large-scale structure and how that affects galactic formation and evolution. In this regard the responses delineated the boundaries of the Cosmic Origins science portfolio and served to underline the broad range of topics and science knowledge that the Cosmic Origins PAG has to consider in their policy deliberations – a breadth that is larger in scope than either the ExoPAG or PhysPAG which are much more focused in their considerations. There are a variety of ways to attempt to pull this kind of information together. The proposals are not taken to be a complete sample from the community, but they are deemed to be representative in terms of the scope and type of science, and the kinds of capabilities they demand – since it is immediately apparent that there are common factors between many of the proposed programs. It is a straightforward thing to assemble a table like Table 1, above, where the specifics of the capabilities called out are tracked and assigned to each and every proposal. A more useful approach is to invert that matrix and make an attempt to evaluate the number of proposed programs that could be enabled by typical technological capabilities that can be envisioned over the next decade.

Following this approach we have chosen to identify certain common foundation-level capabilities that we could see as being pivotal for enabling the science discussed above – these include wavelength coverage, angular resolution, spectral resolution, aperture and field of view.

Addressing the imaging capabilities we can identify certain break points in the scale and size of the capabilities. These are rooted in economies of scale, of emerging technologies and in heritage.

When considering wavelength coverage it is the UV end of the coverage that demands all the attention since the optical and NIR capabilities are relatively straightforward. The first break point is in detector choice around 250nm where silicon response starts to drop dramatically and where one starts to encounter issues with carrier multiplication, which, while not a problem in terms of higher signal per incident photon, does complicate calibration and photometric accuracy. As such below this breakpoint it is advisable to switch to a different detector architecture such as micro-channel plate using a different anode material, or the newly emerging electron multiplying CCD (EMCCD) technology. A second breakpoint is 115nm where the reflectivity of typically-used $MgF_2$-protected aluminum drops dramatically. Such dielectrics protecting aluminum are a commonly-used mirror technology. Below this wavelength a vendor might have to consider use of LiF as an overcoat or the use of material such as SiC, but that is poorly reflective in the visible and therefore counterproductive for the kinds of multiband applications being discussed here. Advances offered by emerging technologies such as atomic layer deposition (ALD) may improve the uniformity and performance of such coatings at thinner layers. The final breakpoint occurs around 92nm where a whole range of issues crop up such as detector architecture as well as prohibitive contamination control during integration and testing.

The angular resolution of an imaging system is rooted fundamentally in both the aperture of the primary optic and the wavelength range to be used – in particular the primary wavelength at which the system is designed to operate and to which such issues as Nyquist sampling will be tuned. Breakpoints in this parameter appear to be more driven by the application itself rather than the technology that drives it – however there are breakpoints in optic aperture that should be considered as well (see below). Revisiting the science drivers listed above, we find that there are relatively coarse angular resolution demands that require little better than the 0.1″ quality or slightly less of the HST-WFPC2 or ACS level of resolution. We characterize that level of resolution with a breakpoint at 50 mas. A second breakpoint occurs at the equivalent of a 4m aperture Nyquist sampled in the mid-optical at a resolution of about 10 mas. Beyond that we get into sub-mas requirements that broach the field of interferometry or truly large apertures in space – which we characterize with a resolution limit of 1 mas.

In terms of actual optical aperture, the drivers here are cost and the capabilities of vendors to fabricate mirrors of the appropriate size, monolith or segmented, based on heritage and existing facilities for integration and test. Because of these factors there are natural breaks on mirror cost. The first of these occurs around 1-2m based on common mirror fabrication factories around the world. These are straightforward mirror sizes to manufacture and can be built as optically "fast" as needed. The next break point comes at Hubble-class sizes, around 2.4m, where we have an immediate economy of scale due to this being a characteristic size adopted by defense vendors for military applications. However, at much larger diameters than this, there is an absence of existing facilities to perform fabrication, polishing, metrology and final integration and end-to-end test, and so this represents a pretty hard breakpoint. The next breakpoint occurs at around 4m, which is driven less by heritage issues (since to our knowledge no 4m mirror has been flown to date) but by the science and capabilities it enables and the present availability of launch vehicle fairings. We see a variety of cases where 4m is the definitive size of aperture that marks the boundary between exoplanet characterization of Jovian-class planets and terrestrial-class planets, between access to comprehensive stellar populations in nearby galaxies and only the upper end of the main sequence, between the ability to find cosmological targets and the ability to characterize them. The final breakpoint occurs at around 8m which is where our current projections for on-orbit mirror sizes tend to diverge when it comes to technological strategic plans – there are few reliable estimates about cost and/or deliverability for this class of primary optic in either monolithic or segmented form, and as such should be regarded as over the horizon for the purposes of defining a mission for the coming decade.

Field of view is affected by aperture, and effective optical system beam speed, but it is most affected by detector real estate. Access to truly large fields of view at the diffraction limit of the optical system feeding that focal plane represents a considerable investment in both development time and resources, but also in bottom line cost and risk associated with delivery of that scale of architecture at the detector level. Recent work on this subject (Scowen et al., 2009) has outlined methodologies that can be adopted to reduce both cost and risk of providing the focal plane arrays, but it will involve the relaxation of some acceptance standards and the development of new calibration and post-processing strategies to deliver an acceptable quality of imaging product from detectors that may not be as cosmetically acceptable as their predecessors. Again the breakpoints are driven by the science demands of the imaging system, but are also rooted in fundamental breaks in cost. We identify small fields of view by a break at 1′ – where the science does not care about field and is generally focused on imaging a single object at a time with the highest fidelity achievable. The next break point occurs at 10′ where the field becomes large enough to allow mapping of modest-sized extended sources and to allow enough resolution to measure entire stellar populations within a particular region. The final breakpoint occurs at 30′ or ½ of a degree where the field size now enables truly large-scale survey work that can cover nearby galactic systems as a whole, or to enable wide-field surveys looking for cosmological or stellar populations whose location is unknown, but for which the angular frequency of the science targets is known to be low and for which a field of this size is necessary to allow a large enough sample to be found to be able to make any kind of statistical statement about their physical nature or properties.

The final metric is that of spectral resolution for dispersive systems. The first breakpoint we identify at $R \approx 1000$, which provides coarse enough resolution to allow modest modeling and measurement of physical properties, but provides enough spectral range on a conventional focal plane detector to allow entire passbands to be acquired in a single observation – the highest of resolutions is neither needed nor affordable for the kind of work envisioned. The next break point occurs at $R \approx 10,000$ where the spectral resolution now allows individual spectral lines to be broken into multiple components across suites of adjacent lines to enable the use of multiple diagnostics for the assessment of physical conditions as well as dynamical information associated with bulk motion in the target objects. The final break point occurs at $R \approx 40,000$ where the fundamental limits of the physical size and scale of the spectroscopic instrument become an issue, where the detector cannot be physically larger than a certain size without violating packaging issues, where the data frame size cannot exceed a determined size without presenting major data handling issues to the mission design, but where the capability now delivers the ability to resolve and split truly daunting blended physical situations in the astronomical targets. It represents a sweet spot between these sets of competing factors in designing spectrographs of this size and scale.

**VI. a. Imaging Performance and Enabling Capabilities**

Using these identified break points we can take the information in Table 1 and use the discussion above to identify which programs are enabled by specific sets of capabilities and contrast that number with the number of programs not enabled by the same, and we present those numbers in Table 2 below. Again, the numbers here are held to be representative and not complete, but we believe them to provide some insight into the kinds of capabilities the astronomical community believes they will need for the next generation science they envision.

Looking at these results from the top down, we see the following results:
- That while all imaging programs may be enabled by pushing the wavelength band down to 92nm, a reasonable compromise can be achieved pushing down to 115nm with conventional imaging architecture and the use of $MgF_2$ over aluminum as a mirror coating material.
- That a reasonable compromise between cost and performance can be struck at an effective angular resolution of around 10mas.
- That while a truly large aperture will enable most programs, a reasonable compromise can be struck with an aperture as low as 2.4m with there being very little gained by moving to 4m of aperture – this can be held as testimony that HST is still a very competitive facility even after 23 years on orbit.
- That an imaging field of view needs to be at least 10′ on a side to enable more than half of the proposed science. This is a good workhorse middle ground with only a few more programs being enabled by a move to a ½ degree field of view.

All this said, and based on the statistics of the enabled programs, we can make the statement that an imaging mission that uses a 2.4m aperture size, has mirrors coated with $MgF_2$ over aluminum, that provides imaging sampling at 10mas, and a combined field of view measured around 10-20′ would enable better than 60% of the proposed science submitted to this opportunity.

### VI. b. Spectroscopic Performance and Enabling Capabilities

In a similar fashion we can study the effect of the break points discussed above on the spectroscopic science enabled. Table 3 presents the results.

In a similar fashion we see the following trends:
- That the majority of spectroscopic science is enabled below a blue-end cutoff of 115nm – delivery of this kind of FUV capability is vital to the kinds of next generation science envisioned by the community.
- That a spectral resolution of at least $R$=10,000 is necessary for the majority of science to be enabled – there is little gained by pushing to truly high resolutions of $R$=40,000 or higher
- That an aperture of at least 2.4m is necessary for the majority of the science – the sweet spot appears to be between 2.4m and 4m in aperture size
- That few programs, relatively, require the use of MOS-like capabilities

In a similar manner to before, we can make the statement that a spectroscopic mission that is between 2.4m and 4m in aperture size, is coated with materials that provide access shortward of 115nm or with few enough reflections to minimize losses, that provides spectral resolution of at least $R$=10,000 would enable better than 60% of the proposed science submitted to this opportunity.

### VII. Conclusions

We counsel the reader not to overinterpret the numbers derived from this analysis, since they are drawn from an incomplete sample, but a sample that nonetheless we contend is representative. The original charge and request made of the community was not bounded by cost or specific technology limitations – we asked for world-class next generation science over the next decade that would be enabled by going to space. The results presented above represent the range of capabilities those science drivers would demand. In this regard we believe that these results define a pretty liberal envelope within which to define UV/visible capabilities that would enable a large fraction (at least two thirds) of the science envisioned. Of course, some science suggested requires capabilities that would have to be planned for over several decades and would represent significant investment and development, but a large fraction could be enabled by relatively modest investment in technologies that already exist today at a low technology readiness level.

In light of these findings, is it possible to envision a single mission that delivers on both the promise of the imaging and spectroscopic capabilities? Combining the findings from above, it is possible to see that a single mission that delivers on the imaging capabilities, but that has a parallel FUV spectroscopic channel where the number of reflections has been limited to two (that may involve the use of an alternate coating material such as LiF or ALD-deposited $MgF_2$) could provide for both sets of design requirements. The spectrograph itself would have to be very high resolution and feed a next generation MCP anode-based detector or EMCCD of the appropriate size and format to deliver the spectral coverage needed. The imaging capabilities demand an aperture of at least 2.4m and a focal plane that is of order 100,000 pixels on a side and delivers optically Nyquist sampling in the mid-optical to NUV. While these are demanding requirements, we do not believe them to be impossible.

The results of this RFI were intended to shape discussion and strategic planning for the next generation of UV/visible missions, and to set priorities for a coherent technology development plan across this decade. In that regard the results are compelling and insightful. We expect these inputs to influence future discussion and mission planning as we approach and prepare for the next Decadal Survey on Astronomy and Astrophysics in 2020.

The efforts to support a future UV/visible mission are certainly not limited to the U.S. and recently ESA has called for science themes and questions for the next two large (L-class) missions as part of the Cosmic Vision 2015-2015 plan. In response to the request of white papers, the European Network for Ultraviolet Astronomy (NUVA) submitted a white paper entitled "European Ultraviolet-Visible Observatory (EUVO", Gómez de Castro et al., 2013), which could lead to future collaborations with the U.S. if this effort is successful.

**Acknowledgements.** We thanks to all the members of the community that freely and openly participated in the different meetings, teleconferences, workshops, and to the many authors and co-authors of the RFI submissions that made possible the description and analysis presented here. In particular, we are grateful to the executive committee of the COPAG for their generous contributions in their efforts of science and technology prioritization for this theme. The Astrophysics Division at NASA Headquarters funded most of the work presented here related to the RFI, activity managed under the overall Cosmic Origins Program.

# VIII. References

Astrophysics Implementation Plan, December 2012, Astrophysics Division, Science Mission Directorate, NASA Headquarters – Revision 1
http://science.nasa.gov/researchers/sara/division-corner/astrophysics-division-corner/

France, K. et al., 2013a, ApJ, 763, 149

France, K. et al., 2013b, Private Communication

Gantner, B. et al., 2011, in UV, X-Ray, and Gamma-Ray Space Instrumentation for Astronomy XVII. Ed.: L. Tsakalakos. Proceedings of the SPIE, Volume 8145, article id. 81450A

Gibson, W.C., et al., 2000, SSRv, 91, 15

Gómez de Castro, A. I. et al. 2013, European Ultraviolet-Visible Observatory (EUVO), White Paper submitted in response to the ESA request for science themes to define the next two large L-class missions to be launched in 2028 and 2034

Green, J., et al., 2012, ApJ, 744, 60

Martin, C.D., 2011, BAAS, vol. 43

McCandliss, S. et al., 2004, in UV and Gamma-Ray Space Telescope Systems. Eds.: Hasinger, G.; Turner, M.J.L. Proceedings of the SPIE, Volume 5488, pp. 709-718

Mendillo, C.B., et al., 2012, in Space Telescopes and Instrumentation 2012: Optical, Infrared, and Millimeter Wave. Proceedings of the SPIE, Volume 8442, article id. 84420E

Milliard, B., et al., 2010, in Space Telescopes and Instrumentation 2010: Ultraviolet to Gamma Ray. Eds.: Arnaud, M.; Murray, S.S.; Takahashi, T. Proceedings of the SPIE, Volume 7732, article id. 773205

New Worlds, New Horizons in Astronomy and Astrophysics, 2010, National Research Council of the National Academies.

Panel Reports, 2010, New Worlds, New Horizons in Astronomy and Astrophysics, National Research Council of the National Academies.

Postman, M., 2009 in Beyond JWST: The Next Steps in UV-Optical-NIR Space Astronomy (editor), Space Telescope Science Institute, March 26-27, 2009.

Redfield, S., 2006, in ASP Conference Series 352, New Horizons in Astronomy, 79.

Scowen, P. et al., 2009, "Large Focal Plane Arrays for Future Missions", submitted to New Worlds, New Horizons in Astronomy and Astrophysics, 2010, National Research Council of the National Academies

Shull, J.M., Smith, B.D., Danforth, C. W., 2012, ApJ, 759, 23

Woodgate, B.E., et al., 1998, PASP, 110, 1183

# IX. Tables

**Table 1**: Summary of RFI Responses and Capability Requirements Provided.

| PI | Investigation Title | Angular Resolution | Telescope Diameter | λ (short) | λ (long) | Field of View | Spectral Resolution | Sensitivity | Photo-metry? | Spectro-scopy? | Spectral multiplexing? | Time domain? | Science Category |
|---|---|---|---|---|---|---|---|---|---|---|---|---|---|
| Gull | How do molecules and dust form in massive interacting winds? | <0.010" | | 3000Å | 7000Å | 2" | 10,000 | <<HST | | Y | MOS | | Stars |
| Provencal | The Importance of White Dwarf Stars as Tests of Stellar Physics and Galactic Evolution | | 2m+ | 912Å | 3000Å | 10'x10' | 50,000 | V~35 | Y | Y | IFU | | Stars |
| Lawler | The Origin of the Elements Heavier than Iron | ~0.1 | | 1900Å | 3050Å | 10'x10' | 60,000 | | | Y | MOS? | | Stars |
| Neiner | UVMag: Stellar physics with UV and visible spectropolarimetry | ? | | 1170Å | 0.87µm | | 25,000 | V~10 | Y; pol | | | Y | Stars |
| Ignace | Importance of time series and polarimetry | | | | | | | | Y; pol | | | Y | Stars |
| Carpenter | Mass Transport Processes and their Roles in the Formation, Structure, and Evolution of Stars and Stellar Systems | <100µ" | 1m x N | 1200Å | 1600Å | 4mas | 10Å | | Y | Y | spectral imaging | Y | Stars |
| Scowen | Understanding Global Galactic Star Formation | 0.020" | 1.5m-4m | 2500Å | 0.95µm | >15'x15' | | | Y | | | | Star Formation |
| Scowen | The Magellanic Clouds Survey - a Bridge to Nearby Galaxies | <0.1" | 2m-4m | 2000Å | ~1µm | 10'x10' | 30,000 | $10^{-16}$ erg/s/cm$^2$/arcsec$^2$ | Y | Y | | | Star Formation; Stars |
| Wofford | Massive Stars: Key to Solving the Cosmic Puzzle | <0.1" | ≥10m | 912Å | 0.9µm | 25" x 25" | 6,000 | | | Y | | | Nearby Galaxies; Stars |
| Barstow | Conditions for Life in the Local Universe | <0.1" | | 1000Å | 3000Å | | 100,000 | | Y | Y | prob N | | Nearby Galaxies; Stars |
| Brown | The History of Star Formation in Galaxies | diff. limit | 8-16m | 3000Å | 9000Å | >= 3' x 3' | | V~35 | Y | | | | Nearby Galaxies |
| Goudfrooij | Near-Field Cosmology and Galaxy Evolution Using Globular Clusters in Nearby Galaxies | .05" | 2m/8m | 2000Å | 5500Å | 20'x20' | 3000 | | Y | Y | MOS x30 | | Nearby Galaxies |
| Williams | The Crucial Role of High Spatial Resolution, High Sensitivity UV Observations to Galaxy Evolution Studies | 0.007" | 8m-10m | 1100Å | 6000Å | 30' x 30' | | | Y | | | | Nearby Galaxies |
| Gordon | A Census of Local Group Ultraviolet Dust Extinction Curves | 0.1" | | 1150Å | 4100Å | | 1000 | | Y | Y | | | Nearby Galaxies |
| Shull | The Baryon Census in a Multiphase Intergalactic Medium | | >4m | <1000Å | | | ~100,000 | 2mÅ EW | | Y | | | IGM |
| Tripp | Quasar Absorption Lines in the Far Ultraviolet: An Untapped Gold Mine for Galaxy Evolution Studies | | | 1000Å | | | like COS | <<HST | | Y | | | IGM |
| Gomez de Castro | Seeking into the anthropic principle - | .01? | | 900Å | 3200Å | 3' | 1000 | | Y | Y | | | IGM |
| Scarlata | The escape fraction of ionizing photons from dwarf galaxies | 1" | | 1000Å | 0.63µm | 10' | 5000 | ~32nd AB | Y | Y | | | IGM |
| Schiminovich | Science from IGM/CGM Emission Mapping | 1-5" | | 1000Å | 4000Å | 4' x 4' / 20' x 20' | 1000-5000 | 5, 100 γ/cm$^2$/s/sr | | Y | IFU / MOS | | IGM |
| McCandliss | Project Lyman: Quantifying 11 Gyrs of Metagalactic Ionizing Background Evolution | ~0.25" | | 950Å | 4000Å | 0.5"$^2$ | few 1000s | $10^{-4}$ FEFU | Y | Y | MOS | | IGM |
| Kriss | Synergistic Astrophysics in the Ultraviolet using Active Galactic Nuclei | 0.01" | 8m | 900Å | 3200Å | 1" | 15,000 – 40,000 | 10 FEFU | | Y | | Y | AGN; IGM |
| Kraemer | Active Galactic Nuclei and their role in Galaxy Formation and Evolution | <100µ" | | 1200Å | 3000Å | 4mas | ~500 | | Y | Y | ?data cube? | | AGN |
| Peterson | UV Spectroscopic Time Domain Studies of Active Galactic Nuclei | ?1" | | 1100Å | 3000Å | | >600 | | | Y | | Y | AGN |
| Hayes | Extragalactic Lyman-alpha Experiments in the Nearby Universe | 0.05" | | 1216Å | 3500Å | 0.1"$^2$ | 100 - 5000 | $10^{-16}$ erg/s/cm$^2$ | Y | Y | Any | | Galaxy Evolution |
| Scowen | Galaxy Assembly and SMBH/AGN-growth from Cosmic Dawn to the End of Reionization | ≤0.040" | 2.4m-4m | 2000Å | ~1µm | 15'x15' | few 100s | ~30th AB | Y | Y | Slitless | | Galaxy Evolution |
| Heap | A UV/Optical/Near-IR Spectroscopic Sky Survey for Understanding Galaxy Evolution | 0.4-0.9" | 0.5m-2.4m | 2000Å | 1.7µm | 0.15"$^2$ | 400-1000 | 0.001 FEFU | | Y | MOS ~400 micro | | Galaxy Evolution |
| Doré | An Optical and Ultraviolet Cosmological Mapper | 30" | 0.5m | 1216Å | 0.85µm | | | $10^{-16}$ erg/s/cm$^2$/arcsec$^2$ | Y | Y | | | Galaxy Evolution |
| Noecker | Exoplanet Science of Nearby Stars on a UV/Visible Astrophysics Mission | diff. limit | 4m | <5000Å | >8000Å | 1" | 100 | | Y; coron. | Y | | Y | Planets |
| Cook | Ultraviolet imaging of exoplanets | | 0.5m-1.5m | | | | | | Y; coron. | Y | | | Planets |
| France | Understanding the Life Cycle of Circumstellar Gas with Ultraviolet Spectroscopy | ~1" | | 912Å, 1200Å | 1800Å, 4000Å | 10'x10' | 150,000 / 3000 | 0.01 FEFU | | Y | MOS | | Planets |
| Wong | Solar System Science Objectives with the Next UV/Optical Space Observatory | 0.05" | | UV | IR | <50" | 2500 | | Y | Y | | Y | Solar System |
| Côté | A Wide-Field, High-Resolution Imaging Space Telescope Operating at UV/Blue Optical Wavelengths | 0.15" | 1m | 1500Å | 5500Å | 0.67"$^2$ | 100-700 | NUV~26 | Y | Y | slitless | Y | Multiple |
| Tumlinson | Unique Astrophysics in the Lyman Ultraviolet | 0.15 | | 912Å | 2000Å? | | 50,000 | ~30th AB | | Y | MOS 10-100x | | Multiple |
| Ulmer | Next Generation Space UV-Vis Space Observatory (NG-SUVO) | | 2.4m | UV | Vis | 6'x6' | | ~10X HST | Y | Y | | | Multiple |

**Table 2**: Effect of Differing Technology Break Points on the Number of Proposed Imaging Science Programs Enabled.

| Parameter | Enabled | Not Enabled |
|---|---|---|
| Waveband: | | |
| ≥ 92nm | 18 | 0 |
| ≥ 115nm | 11 | 5 |
| ≥ 250nm | 4 | 13 |
| Resolution: | | |
| ≥ 1 mas | 13 | 3 |
| ≥ 10 mas | 12 | 4 |
| ≥ 50 mas | 8 | 8 |
| Aperture: | | |
| 1-2m | 7 | 10 |
| 2.4m | 11 | 6 |
| 4m | 12 | 5 |
| 8m+ | 16 | 1 |
| FoV: | | |
| 1 arcmin | 5 | 12 |
| 10 arcmin | 11 | 6 |
| 30 arcmin | 15 | 2 |

**Table 3**: Effect of Differing Technology Break Points on the Number of Proposed Spectroscopic Science Programs Enabled.

| Parameter | Enabled | Not Enabled |
|---|---|---|
| Waveband: | | |
| ≥ 92nm | 22 | 2 |
| ≥ 115nm | 13 | 11 |
| ≥ 250nm | 2 | 22 |
| Spectral Resolution: | | |
| $R$=1000 | 9 | 15 |
| $R$=10,000 | 16 | 8 |
| $R$=40,000 | 18 | 6 |
| Aperture: | | |
| 1-2m | 6 | 18 |
| 2.4m | 12 | 12 |
| 4m | 16 | 8 |
| 8m+ | 20 | 4 |
| MOS: | 8 | N/A |

# X. Figures

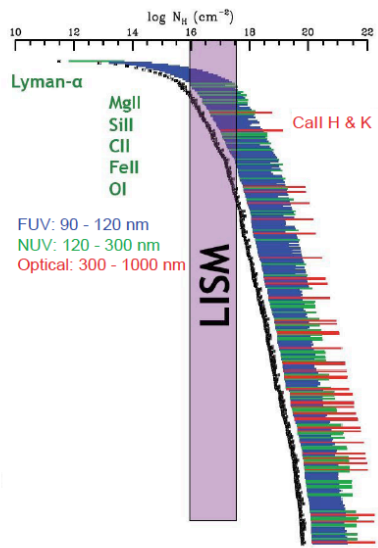

**Figure 1.** All but two of the key diagnostics lines are in the ultraviolet for the local interstellar medium (LISM) (taken from Postman, 2009).